# Large-Scale correlations between QSOs and IRAS galaxies

M. Bartelmann and P. Schneider

Max-Planck-Institut für Astrophysik, Postfach 1523, D–85740 Garching bei München, FRG

**Abstract.** Large-scale structures in the Universe can act as weak lenses and therefore cause an amplification bias on flux-limited counts of background sources provided their luminosity function is sufficiently steep. If galaxies are associated with these structures, as assumed in the biasing hypothesis of galaxy formation, then this weak lensing effect can be detected by searching for correlations between background sources and foreground galaxies on scales of some 10 arcminutes. We have recently investigated this effect theoretically and have shown that highly significant large-scale correlations between 1-Jansky radio quasars and Lick galaxies are present (Bartelmann & Schneider 1993a,b). Here, we extend the analysis, correlating high-redshift 1-Jansky radio quasars with IRAS galaxies. Highly significant correlations are found also in this case; specifically, we find that 1-Jy radio quasars with $z \gtrsim 1.25$ are correlated with IRAS galaxies at the 95% confidence level, which is increased to more than 99% for $z \gtrsim 1.5$. We make use of individual correlation coefficients and demonstrate that they may provide interesting information on individual background sources; for example, this procedure reveals that the 1-Jy source $J2216 - 038$ is weakly lensed by two X-ray bright galaxy clusters in the foreground. A comparison of the sky distribution of those 1-Jy sources which exhibit large correlation coefficients with either Lick or IRAS galaxies with the sky distribution of X-ray bright galaxy clusters from the EMSS tentatively indicates that highly correlated background sources are located preferentially in regions of higher-than-average cluster density. We interpret these results as further evidence for weak lensing by extended structures in the universe which are traced by optical, infrared, and X-ray emission.

## 1 Introduction

In a sequence of papers, we have recently investigated whether large-scale matter inhomogeneities in the universe can cause observable gravitational lensing effects. Based on a simplified, numerically fast scheme of structure formation (Bartelmann & Schneider 1991,1992) and motivated by the claim (Fugmann 1990) that background radio sources appear to be correlated with foreground galaxies on angular scales of some ten arcminutes, we studied whether such correlations could indeed be caused by the lensing effect of large-scale structures alone. We found theoretically that this could indeed be the case, and described a robust and distribution-free statistical test to find these correlations (Bartelmann & Schneider 1993a). A subsequent re-analysis, based on the proposed statistical test, of correlations between optically identified quasars from the 1-Jansky catalogue (Kühr 1981, Stickel 1992, Stickel & Kühr 1993a,b) and galaxies from the Lick catalogue (Shane & Wirtanen 1967, Seldner et al. 1977, Groth 1992) showed that highly significant correlations indeed exist, and that some qualitative trends of these correlations predicted in Bartelmann & Schneider (1993a) could be verified.

However, serious doubts about the validity and reliability of the Lick catalogue have frequently been raised (see e.g., de Lapparent et al. 1986, Geller et al. 1984). It therefore appeared desirable to find further arguments for the existence of the correlations claimed and to repeat the correlation analyses with different, and probably more reliable, galaxy catalogues.

In this paper, we extend the method of analysis by including statistics of rank-order correlation coefficients of individual background sources, and apply these to correlations between 1-Jansky sources and Lick galaxies.

In Sect.2, we describe the selection of galaxies from the Infrared Astronomical Satellite (henceforth IRAS) Faint Source Catalog (hereafter FSC) and the selection of quasars from the 1-Jansky sample. The statistical method is summarized and applied to the data in Sect.3. In Sect.4, we use rank-order correlation coefficients of individual sources and discuss the information contained in their distribution and information which can be obtained from individual highly correlated sources. Finally, Sect.5 contains summary and discussion.

## 2 Sample selection

### 2.1 Galaxies from the IRAS Faint Source Catalogue

We use the IRAS Faint Source Catalogue (hereafter FSC) as the database for positions and fluxes of infrared galaxies. Compared to the IRAS Point Source Catalog (hereafter PSC), the FSC is more sensitive (by a factor of $\simeq 2.5$), but slightly less reliable (to $\gtrsim 94\%$ compared to the near-100% of the PSC). It contains 173044 sources spread over $\simeq 80\%$ of the sky; $\simeq 2\%$ were not covered by the IRAS satellite, and a strip parallel to the galactic equator with galactic latitude $|b| \leq 10^0$ is cut out of the data. IRAS has observed in four spectral bands centered on 12, 25, 60, and 100 $\mu$m, and the respective fluxes are given in the FSC in Jansky for all sources with a detectable signal above some threshold signal-to-noise ratio. With increasing wavelength, the sky is increasingly contaminated by the infrared cirrus emission due to interstellar dust, which makes flux determination less reliable and confusion with nearby sources more probable for increasing wavelength and decreasing galactic latitude.

Those FSC sources with a high- or moderate-quality flux determination at 12, 25, and 60 $\mu$m show a roughly bimodal distribution in a two-colour plane, where colour is defined by the logarithm of shorter-wavelength flux divided by longer-wavelength flux. There is a "bluer" population, identified mainly with stellar sources, and a "redder"population, less numerous than the former, identified mainly with galaxies. Therefore, colour appears as an efficient criterion to select infrared galaxies from the FSC.

Following Strauss et al. (1990), we adopt as a discriminator for galaxies the relation

$$S_{60}^2 \geq S_{12} S_{25} \qquad (1)$$

where $S_n$ is the flux at wavelength $n$ $\mu$m. Due to infrared cirrus and confusion, the sky is only poorly sampled at 60 $\mu$m for $|b| \leq 20^0$.

To avoid poorly determined fluxes, we restrict the sample to $S_{60} \geq 0.3$ Jy, and to exclude strong, nearby sources like the Magellanic Clouds or the star-formation regions in Orion and Taurus, we impose an upper flux limit of $S_{60} \leq 1$ Jy. Table 1 displays the sample sizes after application of the colour criterion and the flux-limit criterion

$$0.3 \text{ Jy} \leq S_{60} \leq 1 \text{ Jy} . \qquad (2)$$

The number distribution of FSC sources as a function of $S_{60}$ is also approximately bimodal, where the two modes are separated by $S_{60} \simeq 0.3$ Jy.

**Table 1.** Sample sizes: numbers of sources in the complete FSC and after application of criteria (1) and (2)

|  | FSC, complete | $0.3 \text{ Jy} \leq S_{60} \leq 1 \text{ Jy}$ | $S_{60}^2 \geq S_{12} S_{25}$ |
|---|---|---|---|
| number of sources | 173044 | 41433 | 35427 |
| fraction | 100% | 23.9% | 20.5% |

By the term "IRAS galaxies" we refer in the following to those $\simeq 35000$ sources in the FSC which satisfy both criteria (1) and (2). Note that this number of sources, spread of $\simeq 80\%$ of the sky, yields a galaxy density of $\simeq 1$ per square degree.

For detailed information on the FSC, the FSC Explanatory Supplement provided by the Infrared Processing and Analysis Center (IPAC) is an excellent reference.

## 2.2 Radio sources from the 1-Jansky sample

As proposed by Fugmann (1990) and in Bartelmann & Schneider (1993b, henceforth BS2), we use here the sample of bright radio sources with a 5-GHz flux above 1 Jy (Kühr 1981). It was communicated to us in the version of May 1992 by M. Stickel and completed by additional optical identifications provided by Stickel & Kühr (1993a,b). In its present form, the 1-Jy catalogue contains 515 sources, 424 or 82.3% of which are optically identified and have spectroscopically determined redshifts. The strip $|b| \leq 10^0$ in galactic longitude is also excluded from the 1-Jansky sample, so that this limitation in the sky coverage of the FSC does not further restrict the selection of 1-Jy sources. Fig.1 shows two charts in aitoff projection of the distribution of IRAS galaxies (bottom panel) and 1-Jy sources (top panel) on the sky in galactic coordinates.

As mentioned above and as is apparent in Fig.1, IRAS galaxies are poorly sampled in the strip $10^0 \leq b \leq 20^0$ in galactic coordinates because of the difficulties in measuring 60-$\mu$m fluxes close to the galactic plane. Therefore, we checked whether our results presented later are changed when we exclude 1-Jy sources with $b \leq 20^0$; we found that this is not the case.

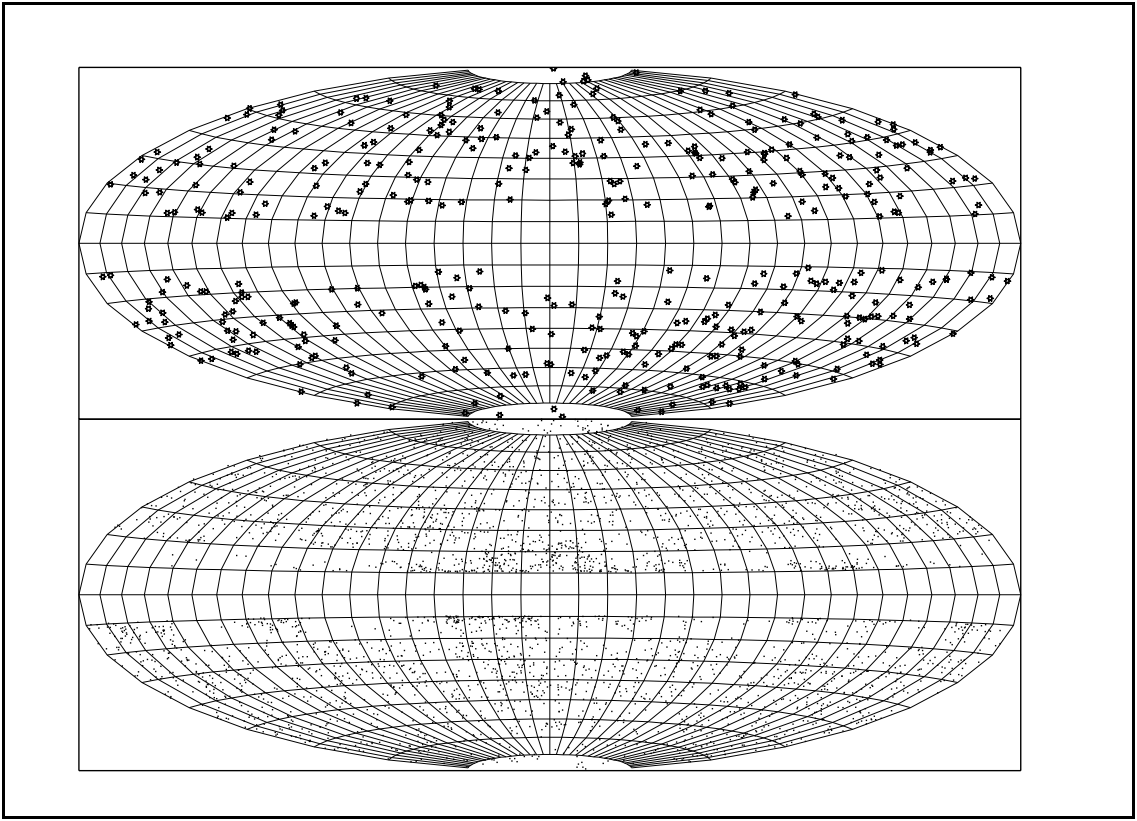

**Fig. 1.** Distribution of 1-Jy sources (top panel) and IRAS galaxies (bottom panel) on the sky; the charts show an aitoff projection of galactic coordinates. (Only one tenth of the IRAS galaxies is plotted)

Most of the optically identified 1-Jy sources with moderate or high redshifts are flat-spectrum QSOs; for instance, there are no sources other than QSOs at redshifts above 1, and of these, 89% show a flat spectrum (spectral index $\geq -0.5$).

We select subsamples from the optically identified 1-Jy sample by imposing a lower redshift limit $z_{\min}$ and an optical magnitude threshold of $m_{\max}$, where $m$ is the visual magnitude.

## 3 Statistical method

We use basically the same statistical method as described in Bartelmann & Schneider (1993a, hereafter BS1) and BS2 to investigate correlations between 1-Jy sources and IRAS galaxies; we thus refer the reader to those papers for details of the procedure. Briefly, the method proceeds in four steps:

1. A symmetric pattern of $n = 25$ equal-area cells is constructed around each 1-Jy source. The number of galaxies in each cell is determined and ranked, yielding a scheme of ranks for each 1-Jy source. These individual rank schemes are averaged over the source subsample (defined by $z_{\min}$ and $m_{\max}$ as mentioned in Sect.2.2), thus yielding one averaged rank scheme per subsample.
2. This averaged rank scheme is again ranked, yielding one scheme of "ranks of mean ranks" per source sample.

3. The distance of the cells from the pattern center is ranked in descending order, i.e., cells closer to the center are ranked higher.
4. Both sequences of ranks, the ranks of the mean ranks of the galaxy counts and the ranks of cell distances, are subjected to Spearman's rank-order correlation test (for a detailed description of this test, see, e.g., Kendall & Stuart 1973). A correlation coefficient $r_{\rm corr} \in [-1, +1]$ is determined from the two sequences of ranks and the number of cells per pattern, and from the (aymptotically) known statistical properties of $r_{\rm corr}$ the significance of the result can be assessed.

To be more specific, let $\mathcal{R}_i^{a,b}$, $i \in \{1, \ldots, n\}$, be the two sequences of ranks. $r_{\rm corr}$ is defined by

$$r_{\rm corr} \equiv 1 - \frac{2D}{D_{\max}},$$

$$D \equiv \sum_{i=1}^{n}(\mathcal{R}_i^a - \mathcal{R}_i^b)^2 \, , \text{ and} \tag{3}$$

$$D_{\max} = \sum_{i=1}^{n} [i - (n - i + 1)]^2 = \frac{n}{3}(n^2 - 1)$$

is the maximum value $D$ can assume. For random ordering, the quantity

$$r_{\rm corr} \sqrt{\frac{n-2}{(1-r_{\rm corr})^2}} \tag{4}$$

follows aymptotically, for $n \gtrsim 10$ in excellent approximation, a Student-t-distribution with $(n-2)$ degrees of freedom, i.e.,

$$dP(r_{\rm corr}) = \frac{(1 - r_{\rm corr})^{(n-4)/2}}{{\rm B}\left(\frac{n-2}{2}, \frac{1}{2}\right)} \, dr_{\rm corr} \, , \tag{5}$$

where ${\rm B}(a, b)$ is the complete beta function.

The hypothesis we have in mind is that 1-Jy sources are correlated with foreground galaxies. The hypothesis tested by Spearman's rank-order correlation test is the opposite; the test yields the probability $\epsilon$ that some correlation coefficient $r_{\rm corr}$ is produced by entirely uncorrelated sequences of ranks. $\epsilon$ is therefore called error probability or error level, since it gives the probability of erroneously adopting the hypothesis of 'correlation'. From Eq.(5), $\epsilon$ is given by

$$\epsilon(r_{\rm corr}) = \int_{r_{\rm corr}}^{1} dP(r_{\rm corr}) \, . \tag{6}$$

For $r_{\rm corr} \to +1$, $\epsilon \to 0$, and for $r_{\rm corr} \to -1$, $\epsilon \to 1$. As seen from Fig.1 of BS1, $r_{\rm corr} = 0.5$ yields $\epsilon = 0.4\%$ for $n = 25$ cells per pattern.

The Lick catalogue is already organized in rectangular cells of $10' \times 10'$ size, and this structure of the catalogue naturally lends itself for constructing cell patterns around each source. Therefore, when analyzing correlations between Lick galaxies and 1-Jy sources, it is only possible to determine the catalogue cell containing the source, which forms the central cell of the pattern. In contrast, the IRAS galaxy sample lists the coordinates of each galaxy. It is then possible to choose concentric ring-shaped cells centered on the 1-Jy source under consideration. As in the case of the Lick catalogue, we choose cells of area 100 (arcmin)$^2$, although a variation of the cell size is possible in principle.

### 3.1 Treatment of ties

When two or more cells of a pattern contain the same number of galaxies, these cells are called 'ties'. It is unclear at first sight how ties should be ranked. Two methods are usually proposed (see, e.g., Kendall & Stuart 1973):

1. Assignment of midranks: all tied cells are assigned the same averaged rank. If, e.g., there are four empty cells in the pattern, they will be assigned rank $(1+4)/2 = 2.5$. In particular, if a pattern was completely empty, each cell would be ranked $(1+n)/2$, or 13 for $n = 25$.
2. Assignment of random ranks: the tied cells are randomly ranked. In the two examples given above, a randomly chosen permutation of the ranks $1\ldots 4$ or $1\ldots n$ would be assigned to the corresponding cells.

Method (1) is easier to apply than method (2). However, when there are many tied cells (in particular, empty cells) in a pattern, application of method (1) seriously corrupts the statistical properties of the derived correlation coefficient. As an example, consider the case of a pattern of $n$ cells, all of which are empty. Following method (1), rank $(n+1)/2$ is assigned to all cells. The correlation coefficient, defined by Eq.(3), is then

$$r_{\text{corr}} = 1 - \frac{6}{n(n^2-1)} \sum_{i=1}^{n} \left[ i - \frac{n+1}{2} \right]^2 = 0.5 \;, \qquad (7)$$

independent of $n$. As mentioned above, $r_{\text{corr}} = 0.5$ with $n = 25$ yields $\epsilon = 0.4\%$, but this 'error level' has nothing to do with the true significance of the result. Therefore, if ties are abundant, method (2) must be applied.

### 3.2 Results for the IRAS galaxy sample

From the number of IRAS galaxies in our sample and the sky coverage of the FSC, we expect roughly one IRAS galaxy per square degree. Since the cell patterns used here have a diameter of some 10 arcminutes, and contain $n = 25$ cells, it is clear that empty cells will be abundant. If the rank schemes are averaged over source samples before computing $r_{\text{corr}}$ and $\epsilon$, the averaged rank scheme will generally contain fewer ties than the individual rank schemes, but due to the very low density of IRAS galaxies, ties will still be frequent also in averaged rank schemes. In contrast to the analysis described in BS1 and BS2, where method (1) was applied, method (2) will therefore be used to treat ties in order to preserve the statistical properties of $r_{\text{corr}}$. As in BS2, we define source subsamples by a lower redshift limit $z_{\min}$ and an upper optical magnitude limit $m_{\max}$. Table 2 displays the results of the rank-order correlation test.

The table shows no significant correlation coefficients for minimum source redshifts $z_{\min} \leq 1$. However, for higher $z_{\min} \geq 1.25$, strong correlations do occur, with significance as high as 99.8% ($\epsilon = 0.2\%$).

At first sight, the optical flux limit $m_{\max}$ has no effect. On the other hand, the source subsamples shrink by roughly a factor of 3 while $m_{\max}$ decreases from 21 to 18, while, at least for $z_{\max} = 1.5$, the strength of the correlation remains approximately constant. This might indicate that the introduction of an optical flux limit partially compensates for the statistical noise due to small subsample sizes, and may therefore tentatively be interpreted as an indication for the multiple-waveband amplification bias (Borgeest et al. 1991).

**Table 2.** Results of the rank-order correlation test between 1-Jy sources and IRAS galaxies for various source subsamples. $z_{\min}$ is the lower bound of source redshifts, $m_{\max}$ is the upper bound of the optical magnitudes of the sources, $N$ is the number of sources in the source subsample, $\mathcal{R}(\langle r_c \rangle)$ is the rank of the mean rank of the central cell, $r_{\text{corr}}$ is the correlation coefficient as defined in Eq.(3), and $\epsilon(r_{\text{corr}})$ is the corresponding error level (see Eq.(6))

| $z_{\min}$ | $m_{\max}$ | $N$ | $\mathcal{R}(\langle r_c \rangle)$ | $r_{\text{corr}}$ | $\epsilon(r_{\text{corr}})$ in % |
|---|---|---|---|---|---|
| 0.50 | 21.00 | 238 | 25.0 | 0.15 | 23.5 |
| 0.50 | 20.00 | 218 | 25.0 | 0.22 | 14.1 |
| 0.50 | 19.00 | 192 | 25.0 | 0.17 | 20.5 |
| 0.50 | 18.75 | 169 | 25.0 | 0.11 | 30.5 |
| 0.50 | 18.50 | 161 | 25.0 | 0.00 | 49.3 |
| 0.50 | 18.25 | 123 | 22.0 | 0.10 | 30.9 |
| 0.50 | 18.00 | 114 | 24.0 | 0.02 | 46.2 |
| 0.75 | 21.00 | 179 | 25.0 | 0.21 | 15.9 |
| 0.75 | 20.00 | 166 | 25.0 | 0.21 | 16.0 |
| 0.75 | 19.00 | 144 | 25.0 | 0.21 | 16.2 |
| 0.75 | 18.75 | 126 | 24.0 | 0.15 | 24.4 |
| 0.75 | 18.50 | 120 | 22.0 | 0.12 | 29.0 |
| 0.75 | 18.25 | 87 | 17.0 | 0.07 | 37.7 |
| 0.75 | 18.00 | 80 | 17.0 | 0.06 | 38.2 |
| 1.00 | 21.00 | 130 | 24.0 | 0.10 | 32.4 |
| 1.00 | 20.00 | 123 | 24.0 | 0.16 | 22.5 |
| 1.00 | 19.00 | 107 | 23.0 | 0.11 | 29.3 |
| 1.00 | 18.75 | 94 | 23.0 | 0.19 | 17.8 |
| 1.00 | 18.50 | 88 | 22.0 | 0.15 | 24.4 |
| 1.00 | 18.25 | 60 | 11.0 | $-0.02$ | 53.8 |
| 1.00 | 18.00 | 54 | 8.0 | $-0.13$ | 73.2 |
| 1.25 | 21.00 | 97 | 24.0 | 0.39 | 2.8 |
| 1.25 | 20.00 | 93 | 24.0 | 0.35 | 4.2 |
| 1.25 | 19.00 | 80 | 22.0 | 0.35 | 4.4 |
| 1.25 | 18.75 | 68 | 22.0 | 0.34 | 4.7 |
| 1.25 | 18.50 | 64 | 23.0 | 0.32 | 6.2 |
| 1.25 | 18.25 | 42 | 10.0 | 0.22 | 14.2 |
| 1.25 | 18.00 | 37 | 10.0 | 0.06 | 38.2 |
| 1.50 | 21.00 | 59 | 19.0 | 0.47 | 0.9 |
| 1.50 | 20.00 | 56 | 17.0 | 0.55 | 0.2 |
| 1.50 | 19.00 | 46 | 19.0 | 0.49 | 0.7 |
| 1.50 | 18.75 | 36 | 23.0 | 0.52 | 0.4 |
| 1.50 | 18.50 | 34 | 19.0 | 0.53 | 0.3 |
| 1.50 | 18.25 | 20 | 20.0 | 0.55 | 0.2 |
| 1.50 | 18.00 | 18 | 21.0 | 0.45 | 1.2 |

The source-redshift limit $z_{\min}$ above which significant correlations occur is considerably larger than the value found by correlating 1-Jy sources with Lick galaxies (BS2). As predicted in BS1, the correlations disappeared for $z_{\min} \gtrsim 1$ because the Lick catalogue is rather shallow in redshift. It is not useful to extend the correlation analysis beyond $z_{\min} \simeq 1.5$ since then the source-subsample sizes become too small.

If we interpret the correlations between 1-Jy sources and IRAS galaxies listed in Table 2 in terms of gravitational lensing by large-scale matter inhomogeneities, the value $z_{\min} \gtrsim 1.25$ of the minimum source redshift for which correlations occur points to lenses at redshifts $\gtrsim 0.25$. For the lensing interpretaion to be valid, a non-negligible fraction of IRAS galaxies should be at that or a higher redshift. In other words, from the correlation results of Table 2 we predict that the sample of IRAS galaxies should be markedly deeper than the Lick catalogue. In the following subsection we will present a simple check for this prediction.

### 3.3 Redshift range of the IRAS galaxy sample

Several redshift surveys of IRAS galaxies were performed [see, e.g., Saunders et al. (1990), Yahil et al. (1991) and references therein]. These redshift surveys are naturally local; the determined redshift distributions peak at radial velocities of a few thousand km/s. Based on these redshift determinations, Saunders at al. and Yahil et al. have derived luminosity functions for IRAS galaxies, and Saunders et al. tested whether there is evidence for a redshift- and/or luminosity evolution of the infrared galaxy population. Based on the assumption that the luminosity function does not change significantly out to redshifts of $z \simeq 0.3$, we can easily obtain a (somewhat crude) estimate of the fraction of distant galaxies in the IRAS sample.

If the comoving number density of IRAS galaxies were constant, their physical number density would be

$$n(z; L) = n_0(L)\,(1+z)^3 \; . \tag{8}$$

It has been discussed in Saunders et al. (1990) that the exponent of $(1+z)$ in Eq.(8) could be as large as $\simeq 7$ due to strong redshift- and/or luminosity evolution. This would enlarge the fraction of moderate-redshift IRAS galaxies compared to the case investigated here. For a conservative estimate, we keep the value 3 for the exponent. In an Einstein-deSitter universe, the volume of a spherical shell at redshift $z$ with width $dz$ is

$$dV(z) = \frac{16\pi}{(1+z)^{9/2}} \left(1 - \frac{1}{\sqrt{1+z}}\right)^2 \, dz \; , \tag{9}$$

so that the number of IRAS galaxies within $dz$ of $z$ is proportional to

$$dN(z) \propto \frac{1}{(1+z)^{3/2}} \left(1 - \frac{1}{\sqrt{1+z}}\right)^2 \, dz \; ; \tag{10}$$

since we will only use relative numbers in the following, the constant of proportionality is irrelevant.

The cumulative luminosity function determined by Yahil et al. (1991) reads, $h_{100}$ being the Hubble constant in units of 100 km/s/Mpc,

$$\begin{aligned}\Psi(l) &= C\,l^{-\alpha}(1+l)^{-\beta} \;, \quad \text{with} \\ \alpha &\simeq 0.5 \;\;,\quad \beta \simeq 1.8 \quad \text{and} \\ l &\equiv \frac{L}{L_*} \;\;,\quad h_{100}^2 L_* \simeq 6 \times 10^9 \, L_\odot \; ; \\ \text{roughly,} &\quad 10^{-2} \lesssim l \lesssim 10^2 \; .\end{aligned} \tag{11}$$

Again, we do not need to know the value of $C$.

Since $S_{60;0} = 0.3$ Jy is the lower flux limit of the IRAS galaxy sample used here, the luminosity of an IRAS galaxy at redshift $z$ required to enter the sample must be greater than

$$l_0 = 4\pi D_L^2(z) \frac{(\nu S)_{60;0}}{L_*} , \qquad (12)$$

where $D_L(z)$ is the luminosity distance and $\nu$ is the frequency. This equation does not take the effects of the K-correction into account, but this neglection is compensated by the neglection of the redshift- and/or luminosity evolution. The fraction of IRAS galaxies visible at redshift $z$ is therefore

$$f(z) = \Psi(l_{\min}) , \quad \text{where}$$
$$l_{\min} = \max(10^{-2}, l_0) . \qquad (13)$$

In that equation, it is implicitly assumed that the fraction of IRAS galaxies with (scaled) luminosities less than $10^{-2}$ or larger than $10^2$ can be neglected.

The number of visible IRAS galaxies with redshift *greater* than $z$ relative to the total number of visible IRAS galaxies is then, from Eqs.(10,13)

$$F(z) \equiv \frac{N(>z)}{N(0)} = \int_z^\infty \frac{dN(z')}{dz'} f(z') \, dz' . \qquad (14)$$

The function $F(z)$ is displayed in Fig.2. The solid line shows $F(z)$ for IRAS galaxies, adopting the luminosity function (11) and a flux threshold of $S_{60;0} = 0.3$ Jy. The short-dashed line shows $F(z)$ for Lick galaxies, adopting a Schechter luminosity function with exponent $\alpha = -1.1$, a magnitude limit of $m_{\rm pg} = 18.9$ mag and the blue K-correction for ellipticals given by Metcalfe et al. (1991).

The figure shows that, based on the assumptions quoted, a non-negligible portion of the IRAS-galaxy sample should extend to redshifts of $0.2\ldots0.4$, while the sample of Lick galaxies shows a steep cutoff at $z \simeq 0.2$. The large deviation between $F(z)$ for IRAS galaxies and Lick galaxies can be attributed to the fact that the luminosity function (11) falls off less steeply than the Schechter luminosity function (algebraically instead of exponentially).

Note that the solid curve in Fig.2 would be raised by taking into account that we also impose an upper flux limit to the IRAS sample. As a result of a much more thorough analysis, Fig.12 of Saunders et al. (1990) also shows that a significant fraction of the IRAS galaxy sample should reach out to redshifts of $\simeq 0.2\ldots0.4$.

### 3.4 Identifications of IRAS galaxies with 1-Jansky sources

One feature apparent in Table 2 is that, for $z_{\min} \leq 1$, the rank of the mean rank of the central cell is almost always 25, while the corresponding correlation coefficients show no significant correlations. For $z_{\min} \geq 1.25$, this trend is reversed: there are no cases with $\mathcal{R}(\langle r_{\rm c}\rangle) = 25$, but the correlation coefficients are large.

First of all, this feature again argues against the use of the rank of the central cell as a statistical measure. We have shown in BS1 and BS2 that cases are frequent where the correlation coefficient is high but $\mathcal{R}(\langle r_{\rm c}\rangle) < 25$, meaning that the rank of the central cell is no reliable estimator for possible correlations. It is seen here that the rank of the central cell can even indicate the opposite trend as the correlation coefficient does.

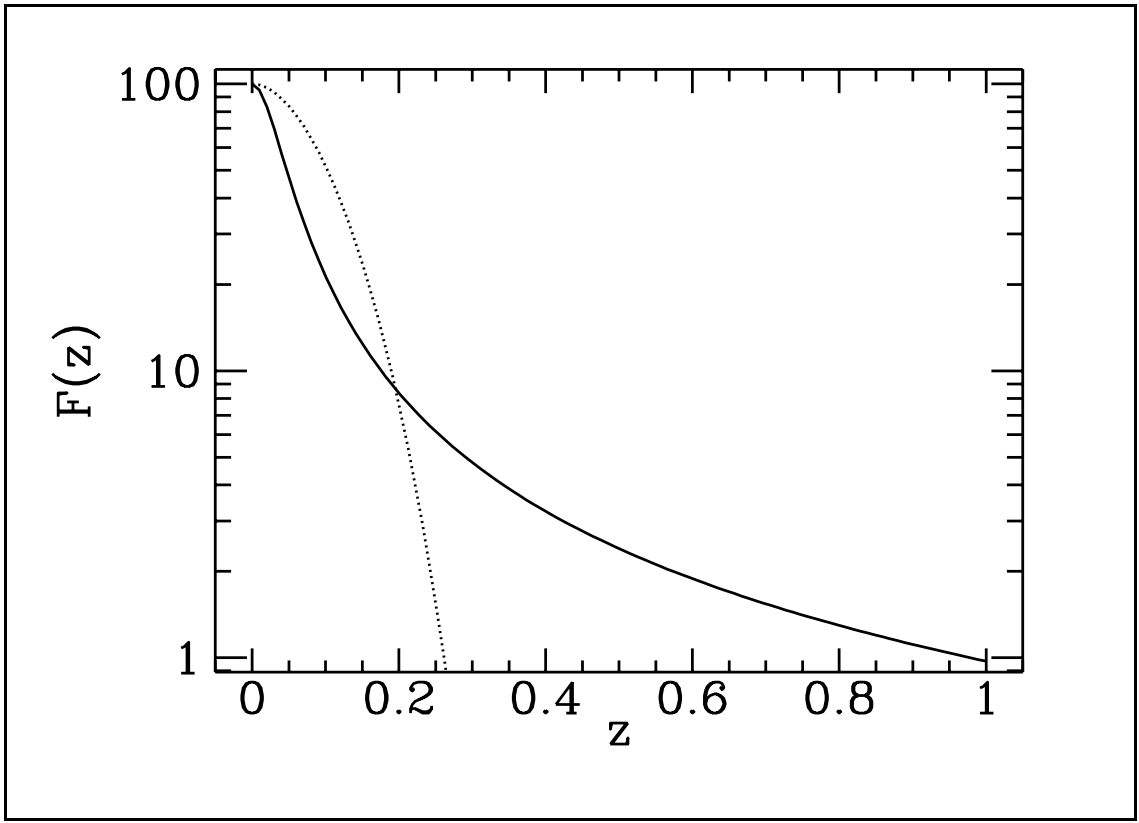

**Fig. 2.** Number of visible galaxies with redshift greater than $z$ relative to the total number of visible galaxies, $F(z)$ from Eq.(14); solid line: $F(z)$ for IRAS galaxies, short-dashed line: $F(z)$ for Lick galaxies

A central rank of 25 indicates that, on average, the number of galaxies in the central cell is highest. Therefore, for small $z_{\min}$, there must be a subset of sources whose cell patterns are dominated by the galaxy counts in the central cell. Isolating 1-Jy sources with very close IRAS galaxies, we found that there is a subset of 1-Jy sources with coordinates deviating from the coordinates of the closest IRAS galaxy by less than ten arcseconds. These sources can tentatively be considered identical with the infrared sources; they are listed in Table 3.

As a test, we have repeated the correlation analysis excluding the central cell from the cell patterns. As expected, the frequent incidence of $\mathcal{R}(\langle r_c \rangle) = 25$ for $z_{\min} \leq 1$ disappears, but the correlation coefficients for $z_{\min} \geq 1.25$ are not changed significantly. We have also checked that an exclusion of the 1-Jy sources listed in Tab.3 from the source subsamples does not change the highly significant correlations either.

## 4 Individual correlation coefficients

Apart from the correlation coefficient of a rank scheme averaged over a subsample of sources, it is also possible to study the correlation coefficients of the rank schemes for individual sources. The number of ties per rank scheme will then be higher on average as compared to the averaged rank scheme, and it is therefore crucial to use method (2) to treat ties.

**Table 3.** 1-Jansky sources with IRAS galaxies in a distance of less than or equal to ten arcseconds. $z$ and $m_\mathrm{v}$ are the redshift and the (visual) magnitude of the optical emission identified with the 1-Jy source, respectively. The 5-GHz flux of the 1-Jy source and the 60-$\mu$m flux of the IRAS galaxy are given in Jy.

| 1-Jy source | IRAS source | $z$ | 5-GHz flux | $S_{60}$ | $m_\mathrm{v}$ |
|---|---|---|---|---|---|
| J0420-014 | F04207-0127 | 0.91 | 1.46 | 0.44 | 17.7 |
| J0537-441 | F05373-4406 | 0.89 | 4.00 | 0.41 | 15.5 |
| J1244-255 | F12440-2531 | 0.63 | 1.36 | 0.40 | 17.4 |
| J1308+326 | F13080+3237 | 0.99 | 1.53 | 0.42 | 19.0 |
| J1406-076 | F14062-0737 | 1.49 | 1.08 | 0.40 | 18.4 |
| J1641+399 | F16413+3954 | 0.59 | 10.8 | 0.60 | 16.0 |
| J1732+389 | F17326+3859 | 0.97 | 1.13 | 0.32 | 19.0 |
| J1803+784 | F18036+7827 | 0.68 | 2.62 | 0.30 | 17.0 |
| J2223-052 | F22231-0512 | 1.40 | 4.51 | 0.59 | 18.4 |

### 4.1 Correlation with Lick galaxies, Kolmogoroff-Smirnov test

First, we study individual correlation coefficients for correlations of Lick galaxies with 1-Jy sources. Although the statistical properties of the Lick catalogue are not well known and probably affected by systematic errors in this catalogue, the Lick sample has the advantage that the number of galaxies per cell is considerably larger than in the case of the relatively rare IRAS galaxies. The analysis proceeds with the same steps described in Sect.3 except that no average over source subsamples is taken.

If the derived correlation coefficients were compatible with no correlation between 1-Jy sources and foreground galaxies, they should be drawn from the Student-t-distribution of Eq.(5). Significant deviations of the distribution of correlation coefficients from the distribution (5) can be quantified with a Kolmogoroff-Smirnov (henceforth KS) test. Fig.3 shows histograms for four different values of $z_\mathrm{min}$ of the correlation coefficients for individual Lick sources as compared to the histogram expected from Eq.(5).

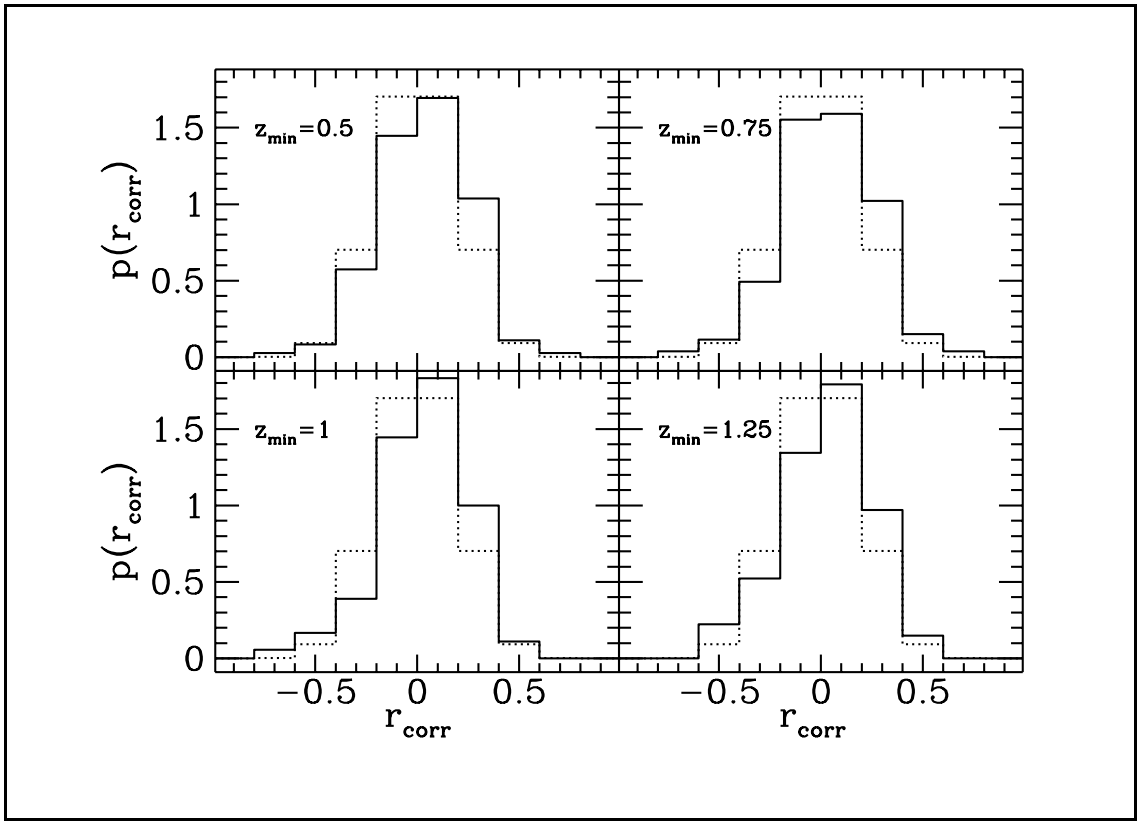

**Fig. 3.** Histograms for correlation coefficients of individual 1-Jansky sources with Lick galaxies (solid line) compared to the corresponding histogram expected from Eq.(5) (short-dashed line). The panels show results for $z_{\min} \in \{0.5, 0.75, 1.0, 1.25\}$ (top left, top right, bottom left, bottom right, respectively). A slight asymmetry of the measured histograms towards positive correlation coefficients is apparent for all four panels

From the figure, a slight shift towards positive correlation coefficients of the measured distributions is apparent. Fig.4 quantifies this appearance using a KS test.

Briefly, the KS test tests whether two distributions can be considered identical. Its result, the significance level $\epsilon_{\rm KS}$, can be interpreted as the error probability for a rejection of the null hypothesis of identical distributions. Fig.4 shows that, for low values of $z_{\min}$, the measured and the expected distributions of $r_{\rm corr}$ are *not* identical with an error probability of a few percent, while for increasing $z_{\min}$, no significant deviations between the two distributions are found. This result is in qualitative agreement with the results from considering correlation coefficients for subsamples of 1-Jy sources (see BS2): for low $z_{\min}$, the significance for correlations is high, and it vanishes for increasing minimum source redshifts.

It is clear that the KS test based on the distribution of individual rank-order correlation coefficients yields less significant results than the rank-order correlation test performed with rank schemes averaged over source subsamples. By the treatment of ties according to method (2), the distribution of $r_{\rm corr}$ is pushed towards the distribution (5), making the distributions similar to each other. Moreover, the KS test compares cumulative distributions, so that a few high values of $r_{\rm corr}$ are 'smoothed out' and thus dominated by the majority of cases where $r_{\rm corr} \simeq 0$. In contrast, the rank-order test

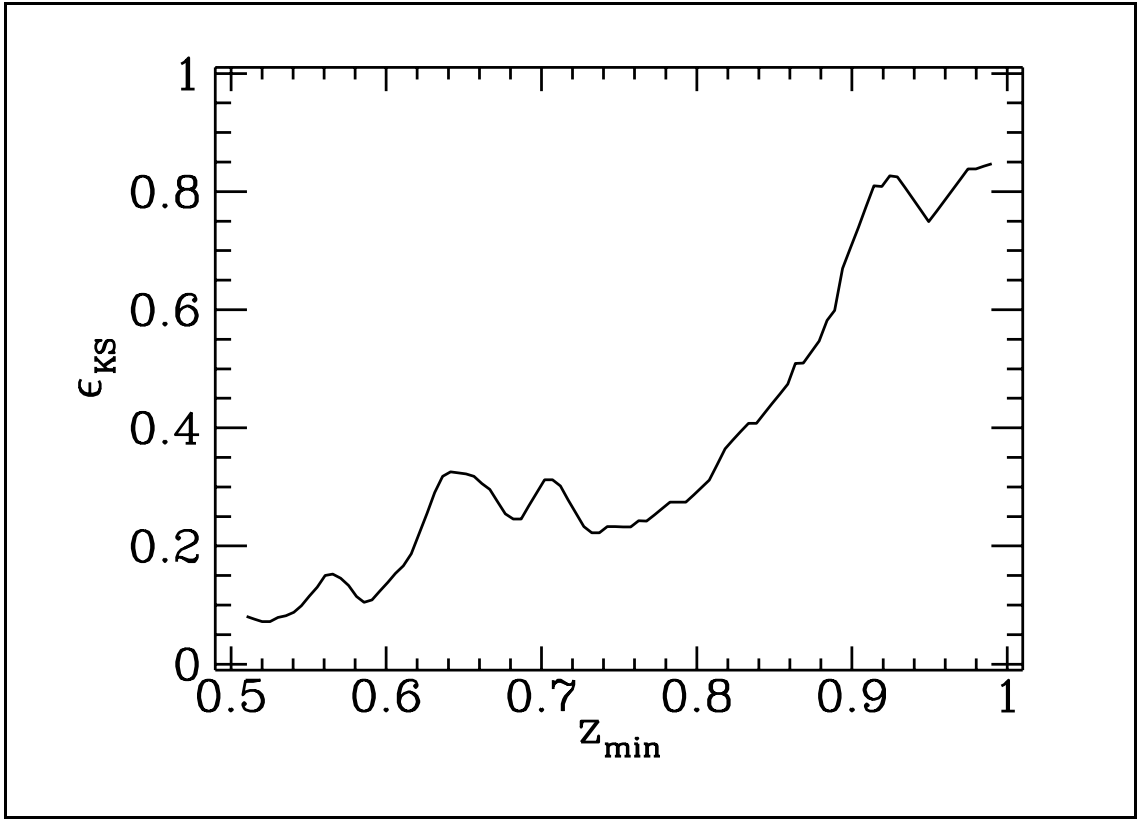

**Fig. 4.** Significance level $\epsilon_{KS}(z_{min})$ from a KS comparison of the distribution of $r_{corr}$ for 1-Jy sources with the distribution (5) expected for no correlation between sources and Lick galaxies. $\epsilon_{KS}$ gives the significance for the two distributions to be identical. Therefore, low values of $\epsilon_{KS}$ indicate that the hypothesis of the two distributions to be identical can be rejected with an error probability of $\epsilon_{KS}$

preserves the information contained in such rank-schemes which yield high correlation coefficients even if they are averaged with a larger number of rank schemes exhibiting no significant correlation.

For comparison, Fig.5 displays ten pairs of histograms similar to those of Fig.3, obtained with the same method of analysis, but by randomly assigning positions to 1-Jy sources. The slight positive asymmetry of the measured histograms as compared to the expected distributions seen in Fig.3 disappears on average, and a KS test also yields high significance for the null hypothesis of no correlations.

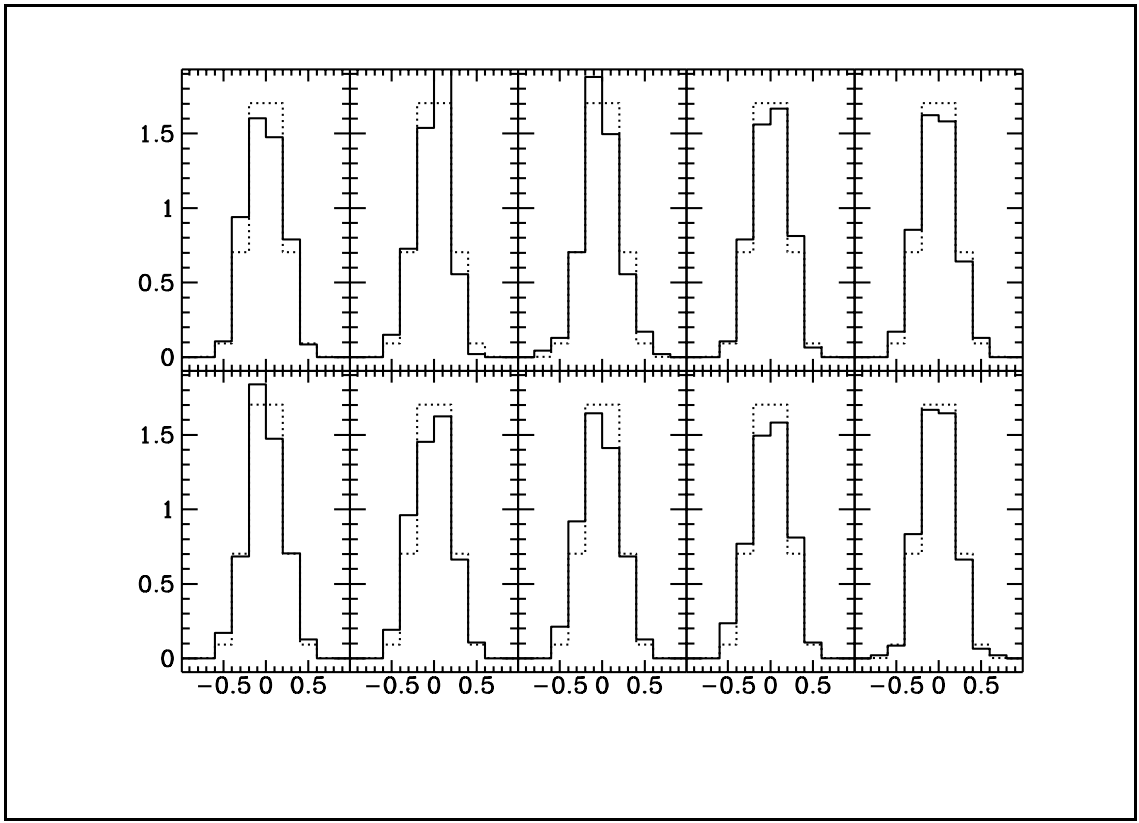

**Fig. 5.** Ten pairs of histograms similar to those of Fig.3, however obtained after randomly assigning positions to 1-Jy sources. It is seen that on average positive deviations of the measured histograms from the expected ones are as frequent as negative deviations. A KS test also shows high significance for no correlations between those randomly positioned 1-Jy sources and Lick galaxies

It is instructive to investigate closer those 1-Jy sources which show the largest individual correlation coefficients. As an example, we have looked for X-ray bright galaxy clusters from the Einstein Medium Sensitivity Survey (hereafter EMSS) in the vicinity of those highly correlated sources. The outstanding case is that of $J2216-038$, $z = 0.901$, which yields the maximum correlation coefficient, $r_{\rm corr} = 0.66$, giving $\epsilon(r_{\rm corr}) \simeq 0.1\%$. Close to that source, the *two* X-ray bright clusters $MS2216.0-040$ and $MS2215.7-040$ are found. [Note that these clusters were only included in the EMSS *because* this 1-Jy radio quasar is there.] Table 4 summarizes the relevant data of the two clusters.

**Table 4.** Some data of the two X-ray bright clusters located close to the 1-Jy source $J2216-038$

| Cluster name | $z_{\rm cluster}$ | X-ray luminosity in $10^{44}$ ergs/s |
|---|---|---|
| $MS2216.0-040$ | 0.09 | 1.58 |
| $MS2215.7-040$ | 0.09 | 1.26 |

The X-ray luminosity of both clusters indicates large masses, and their redshift is roughly one tenth of that of the source, showing clearly that they are foreground objects. The low redshift of the clusters also enables a larger fraction of their galaxies to enter the

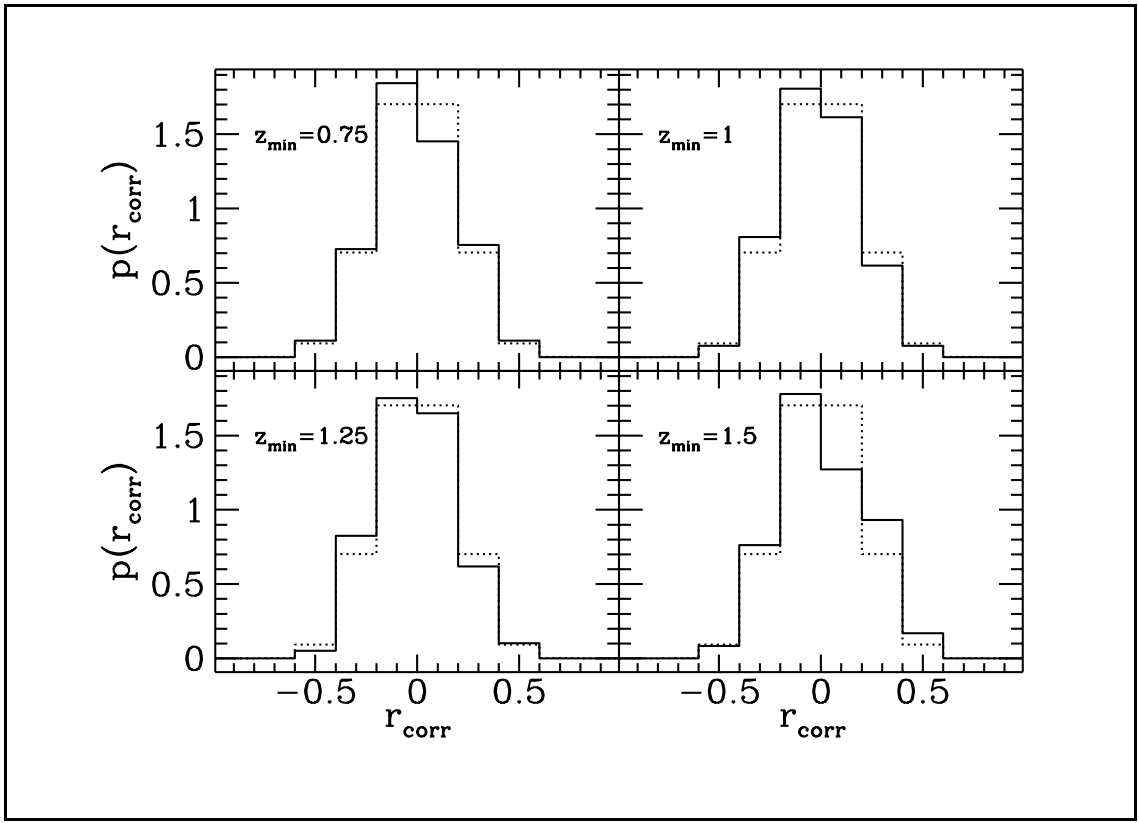

**Fig. 6.** Histograms similar to those of Fig.3, but obtained from correlating 1-Jy sources with IRAS galaxies

Lick catalogue. The separation between the 1-Jy source and the clusters is too large for strong lensing, but a crude estimate of the lensing effect, based on a velocity dispersion of 1000 km/s for both clusters, yields that the source should be magnified by a factor of $\simeq 1.5$. The incidence of two clusters in the foreground of a 1-Jy source demonstrates the capability of the proposed rank-order statistical method to pick out weakly lensed sources which are, due to the magnification by the lenses, correlated with the galaxies associated with the lenses. However, the quoted case of J2216 − 038 is the most spectacular one among those 1-Jy sources showing high correlation coefficients. There are EMSS clusters in the neighborhood of other such sources too, but their separation is too large to expect any significant amount of lensing from them individually. See Figs.7 and 8 for further information.

## 4.2 Correlation with IRAS galaxies

The distribution of individual correlation coefficients obtained from correlating 1-Jy sources with IRAS galaxies are displayed in Fig.6. From the figure, there is little indication for significant deviations between the measured and the expected histograms; only in the case $z_{\min} = 1.5$ (bottom right panel) a slight positive asymmetry of the measured histogram is apparent. A KS test for deviations between the two distributions is negative, i.e., the KS test cannot significantly distinguish between the measured and the expected distributions.

One reason for the absence of significant deviations between the measured and the expected distributions of individual rank-order correlation coefficients in the case of IRAS galaxies is that the KS test is an asymptotic test. This means that the statistical behaviour of the quantity tested is only known exactly in the limit of infinitely many data points. For the correlations with IRAS galaxies, we expect from Table 2 significant correlations for $z_{\min} \gtrsim 1.25$. For such limiting redshifts, the source subsamples become small, and the asymptotic range of the KS test is left.

Another and more important reason is that the number density of IRAS galaxies is small. This causes the majority of cells to be empty and therefore increases the number of tied cells. As discussed in Sect.3, tied cells must be assigned random ranks if they are frequent. Therefore, the distribution of individual correlation coefficients is dominated by the large number of tied, and therefore randomly ranked, cells which makes the distribution indistinguishable from the theoretical distribution of Eq.(5).

### 4.3 Distribution of highly correlated sources on the sky

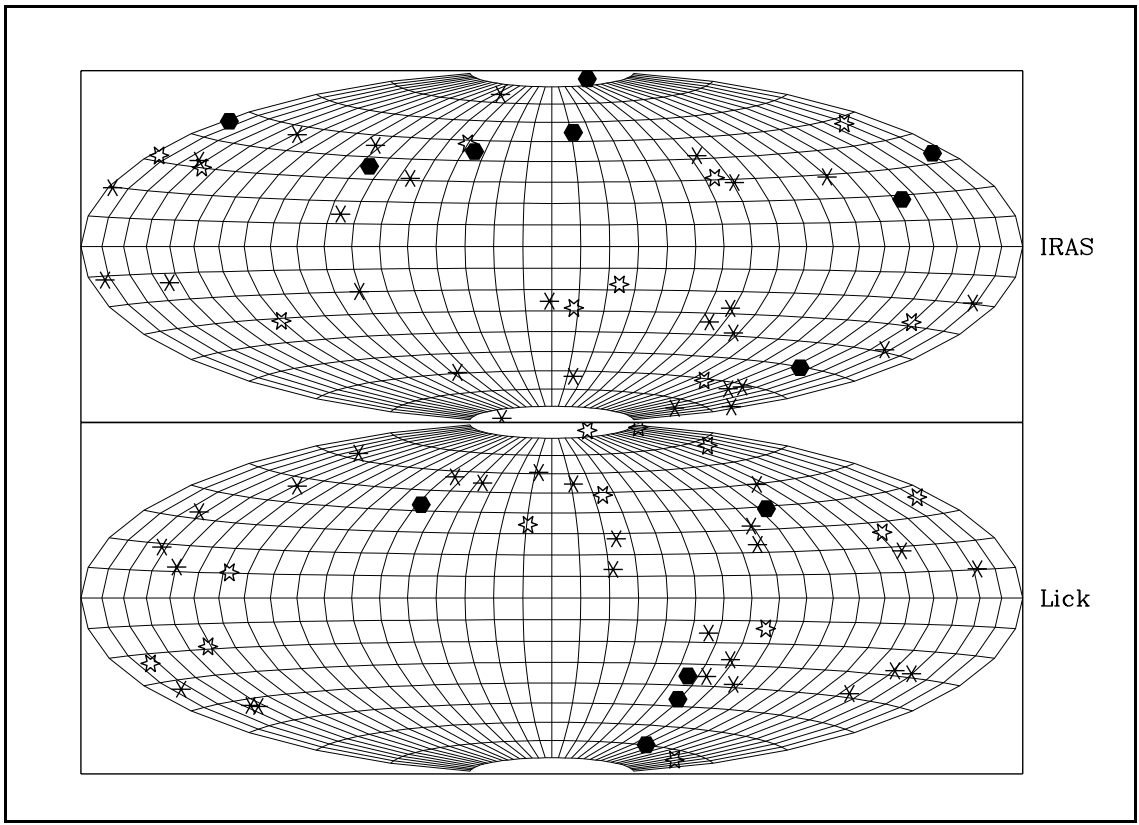

**Fig. 7.** Distribution of such 1-Jy sources on the sky (aitoff map in galactic coordinates) which have high individual correlation coefficients with Lick galaxies (bottom panel) and with IRAS galaxies (top panel). The asterisks denote sources with $0.2 \leq r_{\mathrm{corr}} \leq 0.3$, the stars sources with $0.3 \leq r_{\mathrm{corr}} \leq 0.4$, and the filled hexagons sources with $0.4 \leq r_{\mathrm{corr}}$

The two preceding subsections have shown that the additional *statistical* information obtained from the distribution of individual rank-order correlation coefficients via the KS test is less significant than the correlation coefficients obtained from averaged rank schemes in the case of Lick galaxies and not at all significant in the case of IRAS galaxies due to the large number of tied cells in the latter case. However, the example of the source J2216−038 has demonstrated that the individual correlation coefficients may be useful to obtain some information on individual sources. In particular, the distribution of highly-correlated sources on the sky might contain information about the location of possible dark and extended lenses. Fig.7 shows the distribution of those 1-Jy sources on the sky which have large individual correlation coefficients with Lick galaxies (bottom panel) and IRAS galaxies (top panel). The asterisks denote sources with $0.2 \leq r_{\rm corr} \leq 0.3$, the stars sources with $0.3 \leq r_{\rm corr} \leq 0.4$, and the filled hexagons sources with $0.4 \leq r_{\rm corr}$.

Note that all the filled hexagons in Fig.7 are well away from the galactic plane. For comparison, Fig.8 shows a chart of the positions of galaxy clusters from the EMSS on the sky.

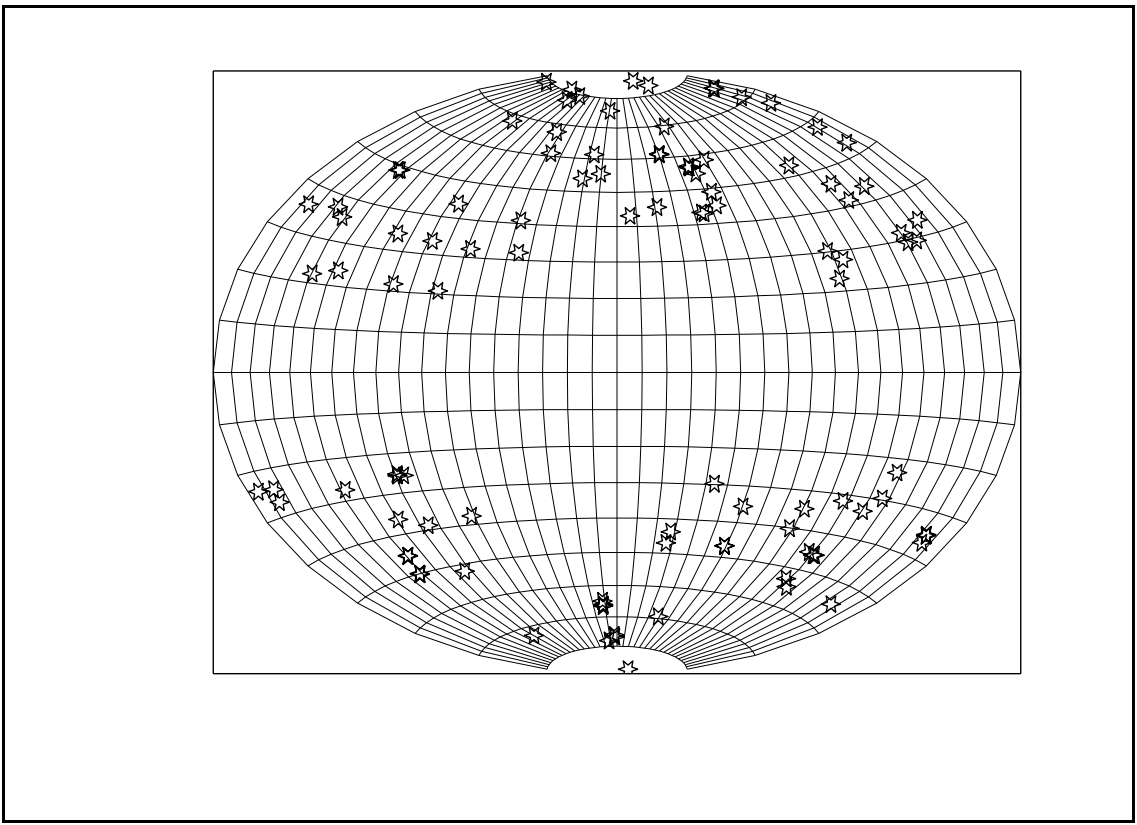

**Fig. 8.** Positions of EMSS clusters on the sky in galactic coordinates, to be compared with the two panels of Fig.7

From a comparison of Figs.7 and 8, it appears that those 1-Jy sources which have high correlations coefficients with either Lick or IRAS galaxies are preferentially located in those regions where the number density of EMSS clusters is also higher than average; in particular, note the region around $l \simeq 60^0$, $b \simeq -40^0$ and around the galactic north pole. Although this inspection by eye is not sufficient to draw any significant conclusions, it might indicate that regions with a higher-than-average density of X-ray bright clusters coincide with regions containing highly-correlated 1-Jy sources. Again, the X-ray bright clusters – with the exception of the cases listed in Tab.4 – are too distant from the 1-Jy sources to individually act as strong lenses, but they probably trace more extended structures which may act as weak lenses.

## 5 Summary and discussion

We have analyzed correlations between moderate- and high-redshift 1-Jansky radio sources with IRAS galaxies from the FSC, using the rank-order correlation test introduced in BS1 and applied to the case of Lick galaxies in BS2. We find (see Table 2) that there exist highly significant correlations on scales of some 10 arcminutes for 1-Jy sources at redshifts $z_{\min} \gtrsim 1.25$.

These correlations may be due to lensing by extended, mostly dark, structures in the universe on scales of 5 to 10 Mpc. Although lensing by such structures is truly weak, it can account for a noticible amplification bias when the source luminosity function is sufficiently steep (see BS1 for a theoretical investigation). The correlations with foreground objects like IRAS galaxies are then caused by the association of galaxies with regions of enhanced matter density as expected in the biasing scenario of galaxy formation. Therefore, such correlations, if they are indeed due to lensing, can serve as the first direct proof of the biasing hypothesis.

The correlations found and displayed in Table 2 occur at higher source redshifts than in the case of correlations of 1-Jy sources with Lick galaxies investigated in BS2. This fact finds a natural explanation within the lensing hypothesis if the redshift range of the IRAS galaxy sample is larger than that of the Lick catalogue. Arguments for this being indeed the case have been collected in Sect.3.3 and displayed in Fig.2; see also Saunders et al. (1990). In Sect.4, the probably conservative assumption was made that the comoving number density of IRAS galaxies is constant, neglecting a (possibly strong) luminosity evolution, which would further expand the redshift range of the IRAS sample.

We therefore propose to interpret the correlations found in terms of gravitational lensing by large-scale structures. There is no obvious reason why infrared galaxies should be physically associated with bright radio sources at redshifts above 1. We have shown in Sect.3.4 – see Table 3 – that some low-redshift 1-Jy sources are probably identical with the IRAS galaxies whose positions agree within better than ten arcseconds, but while this causes the rank of the mean rank of the central cell to be high (24 or 25), it does not cause high correlation coefficients which are formed using the complete information contained in the cell pattern. Moreover, the probably physical associations between IRAS galaxies and 1-Jy sources at low redshifts disappear when the source redshift is limited to $z \gtrsim 1$, which indicates that the infrared emission possibly associated with those sources may be too faint to have been detected by IRAS.

We have argued in BS2 that the multiple-waveband amplification bias (Borgeest et al. 1991) is probably apparent from the fact that an optical flux limit imposed in addition to the radio flux limit on the 1-Jy sample decreases the source subsample sizes, but nevertheless increases the correlation strength. The opposite would be expected without the multiple-waveband bias because of increasing statistical noise. This effect is less clear from the data collected in Tab.2. However, it may be interpreted as indirect indication for the multiple-waveband bias that, although imposing an optical flux limit decreases the size of the source subsamples and should therefore enhance statistical noise, the correlation strength remains roughly constant for $z_{\min} \geq 1.25$.

Additional information can be drawn from the distribution of correlation coefficients for individual sources. We have investigated the distributions of $r_{\mathrm{corr}}$ for correlations between 1-Jy sources and Lick galaxies as well as with IRAS galaxies. Using a Kolmogoroff-Smirnov test to compare the measured distribution and the distribution expected for the case of no correlations between galaxies and 1-Jy sources, we have shown that, in the case of Lick galaxies, the hypothesis of no correlations can be rejected at high significance for small $z_{\min}$, while for high source redshifts, the KS test cannot distinguish between the 'random' and the measured distributions. This is in qualitative agreement with the results of BS2. However, as we discussed in Sect.4, the treatment of ties appropriate for individual rank-order correlation coefficients causes the KS test to be less significant than the rank-order correlation test. This fact, together with the frequent occurrence of tied cells in the case of IRAS galaxies, is the reason that the KS test cannot distinguish between the 'random' and the measured distributions of $r_{\mathrm{corr}}$ in the latter case.

While it is therefore not possible to extract statistical information about the correlations claimed beyond the results of the rank-order test, the case of the 1-Jy source J2216 − 038 has demonstrated that at least in particular cases the individual rank-order correlation coefficients may be used to isolate incidences of weak lensing; for J2216 − 038, two X-ray bright clusters were identified as lensing agents. Sky maps of those 1-Jy sources which exhibit high correlation coefficients with both Lick and IRAS galaxies (see Fig.7), compared to a sky map of the X-ray bright clusters from the EMSS survey, tentatively indicate that highly correlated 1-Jy sources are overabundant in regions where also the density of EMSS clusters is higher than average.

We therefore conclude that the analysis of correlations between high-redshift 1-Jy sources with IRAS galaxies presented here further strengthens the evidence for weak lensing by extended structures in the universe, which are detected by their associated galaxies, which in turn would argue for the biasing hypothesis of galaxy formation. Moreover, the combination of information obtained in wavebands as different as the optical (Lick galaxies, counted in the photographic band), the infrared (IRAS galaxies, peak emission around 60 $\mu$m), and the X-ray band of the Einstein satellite (EMSS clusters) makes explanations for the correlations found other than that based on the lensing hypothesis even harder.

*Acknowledgements.* We want to thank all those persons and institutions who provided data – together with many helpful informations about these data – for this kind of analysis. In particular, we are indebted to Ed Groth for the Lick catalogue, to Manfred Stickel for the 1-Jy sample with its various updates, to the Infrared Processing and Analysis Center (IPAC) for the FSC Explanatory Supplement, the NASA Data Archive and Distribution Service (NDADS) for the FSC itself, and the Einstein On-line Service, Smithsonian Astrophysical Observatory, for the EMSS catalogue. In addition, Emilio Falco and Marc Davis provided many very useful comments.